\documentclass[USenglish,twocolumn]{article}

\usepackage[utf8]{inputenc}
\usepackage[big]{dgruyter}
 
\usepackage{graphicx}
\usepackage{txfonts}
\usepackage{isotope}
\usepackage{color}
\usepackage{hyperref}
\usepackage{setspace}
\usepackage{multirow}

\def\mnras{MNRAS}
\def\apjl{ApJL}              
\def\aap{A\&A}

\def\nature{Nature}
  
\begin{document}

\articletype{Research Article{\hfill}Open Access}

\author*[1]{P\'eter N\'emeth}


\affil[1]{Astroserver.org, 8533 Malomsok, Hungary, E-mail: peter.nemeth@astroserver.org}

\title{\huge Astroserver -- Research Services in the Stellar Webshop}

\runningtitle{Astroserver.org}


\begin{abstract}
{A quick look at research and development in astronomy shows that we live in exciting times.
Exoplanetary systems, supernovae, and merging binary black holes were far out of reach for observers two decades ago and now such phenomena are recorded routinely. 
This quick development would not have been possible without the ability for researchers to be connected, to think globally and to be mobile. 
Classical short-term positions are not always suitable to support these conditions and freelancing may be a viable alternative. 
We introduce the Astroserver framework, which is a new freelancing platform for scientists, and demonstrate through examples how it contributed to some recent projects related to hot subdwarf stars and binaries. 
These contributions, which included spectroscopic data mining, computing services and observing services, as well as artwork, allowed a deeper look into the investigated systems. 
The work on composite spectra binaries provided new details for the hypervelocity wide subdwarf binary PB\,3877 and found diverse and rare systems with sub-giant companions in high-resolution spectroscopic surveys. 
The models for the peculiar abundance pattern of the evolved compact star LP\,40-365 showed it to be a bound hypervelocity remnant of a supernova Iax event.
Some of these works also included data visualizations to help presenting the new results. 
Such services may be of interest for many researchers.}
\end{abstract}

\journalname{Open Astronomy}
\DOI{DOI}
  \startpage{1}
  \received{..}
  \revised{..}
  \accepted{..}

  \journalyear{2017}
  \journalvolume{1}

\maketitle

\section{Introduction}

\subsection{The role of freelancing}

Astronomy has always been a team game and future challenges in research will likely require ever larger collaborations. 
With the sophistication of research instruments, data reduction and analysis methods there is a gradual increase in the degree of specialization, which requires collaboration among many individuals.
For an extreme example it is enough to mention gravitational-wave astronomy, the new and emerging branch of observational astronomy with hundreds of collaborators. 
Large surveys, multi-object spectroscopy and massive transient search programmes are other such examples.
Forefront research and development now require a synergy among many researchers from several fields with diverse expertise, and a similar trend is observed in the presentation of results. 
From drafting publications to the final data visualizations a variety of tools, skills and assets are needed.

All of these concerns currently in research astronomy present many challenges to researchers. 
These challenges are not only science problems; there are some on the management side too.
The problem is, that research astronomy as a full-time tenure position often does not work out and sometimes proves to be an illusion.
It must be associated with a real full-time job, like: professorship, lecturing, IT work, administration, instrumentation, etc. 
After all that, it is always the passion that keeps research alive, as well as the sacrifice of researchers who are willing to take overtime to make the next step at the frontiers of science. 

The peak of research output has shifted in recent years to graduate and post-graduate studies. 
Sadly, but quite understandably, after graduate school many who are otherwise talented and well-rounded for research, but not passionate to pursue astronomy without a predictable pay-off, get discouraged \citep{nature17}. 
With limited postdoctoral research positions, both in length and number, there is a constant struggle to find the next workplace and go through the associated regular relocations. 
Still, postdoctoral appointments are extremely important to improve the quality of researchers and provide the mobility this personal development requires. 
However, after all these, unfortunately too often, many qualified researchers with years of postdoctoral experience, again, simply drop out or return to hobby astronomy. 
The decreasing number of tenure and postdoctoral positions also point in this direction. 
This track in the long run is inefficient and may prove even fateful for the future of astronomy (or science in general). 
In the end everybody is overbooked with extra work and are less able to do research, the quantity of which is therefore decreasing.

Is this the direction astronomy is headed? -- It is also possible that we just see a shift in research methodology, administration and management.
These issues are not unique to astronomy. 
Many research areas have seen an increase in freelancing. 
Indeed, the flexible hours and the opportunity to work from home together with the information technologies of the 21st century make freelancing a tempting alternative to classical positions. 
Certainly, freelancing has its own risks and will never replace academic positions. 
Actually, competitive freelancing relies on the foundations of academic research and complements it.
Some tasks can be more efficiently done in this framework than in academic research.
A good example of this is model atmosphere work and the associated code development. 
These tasks are so specialized and complicated that major improvements often require years of work, if not a decade.

There are many short term projects in which qualified researchers are invaluable, but a position cannot be opened either due to a limitation in funding or the short time-line of the given project. 
Master student research projects are limited to only a few semesters, which allows relatively quick analyses and results. 
PhD projects are longer, however they focus on a specific problem in great detail. 
Often such specific research projects depend on boundary conditions or input parameters taken from other fields, for which collaborations are necessary. 
In these situations the help of occasional postdoctoral researchers may be handy and well justified. 

Astroserver was established to promote freelancing and connect researchers in diverse fields to form collaborations for the good of astronomy. 
Many of the services can be part of larger research projects. 
Here is a list of a few of such tasks: 
\begin{itemize}
\item Content production: Media, image, animation, daily news or technical text, and web content. 
\item Internal reviews, proposal and observation planning, and feasibility evaluations.
\item Development: Implementation of new algorithms or applications.
\item Specialized or legacy codes: Support for code that works, but only a handful of experts can run. 
\end{itemize}
In many cases it may be faster and cheaper if such tasks are outsourced, rather than hiring someone to a classical position. 

Section \ref{maids} briefly describes the computer infrastructure we use for atmosphere modeling work. 
In Section \ref{services} we present a general overview of the services currently offered or targeted by Astroserver. 
Section \ref{contrib} demonstrates some recent research projects with Astroserver contributions and finally we summarize the key points in Section \ref{summary}.


\subsection{MAIDS -- Model Atmosphere Investigations and Data Server\label{maids}}
Scientific research, either theoretical or observational, inevitably requires access to a suitable infrastructure. 
For this reason Astroserver maintains a custom-made computer cluster designed for CPU intensive sequential codes, such as the non-LTE model atmosphere code {\sc Tlusty} (\citealt{hubeny17}). 
The nodes of this cluster are named after women in honor of  the "Harvard Computers" and their pioneering work in stellar spectral classification \citep{welther10}. 
Equipped with the best single thread performance CPUs currently available on the market, MAIDS is an optimal platform for projects where a small number of sequential models and simulations are calculated in an iterative procedure, such as the $\chi^2$ minimization simplex or steepest-gradient fitting methods.
The current version of the cluster delivers 113 GFLOP/s CPU performance and it is easily scalable on demand. 
Although industrial servers provide a larger number of CPU nodes, such tasks with {\sc Tlusty} models would be less efficient on those, due to their lower single thread performance. 
For the very same reason, GPU computing with sequential codes is not viable either. 
Attempts have been made to convert {\sc Tlusty} (the mathematical problem) to a parallelised version, but a production version of the code requires future work.  
However, even though the model atmosphere code itself is sequential, the fitting procedure can make use of multi-processing and, where possible, the models required for a single iteration should be calculated in parallel.

\section{Services\label{services}}
\subsection{Stellar spectroscopy with {\sc XTgrid}}
One of the main research areas of Astroserver is stellar spectroscopy and multi-wavelength spectral analyses of (hot) stars and binaries. 
Atmospheric parameters are derived for (apparently) single stars and composite spectrum binaries using the non-Local Thermodynamic Equilibrium (non-LTE) model atmosphere code {\sc Tlusty} together with the {\sc XTgrid} fitting procedure \citep{nemeth12}. 
The steepest-descent standard $\chi^2$ minimization fitting procedure adjusts the atmospheric parameters and the properties of the binary members until the observed spectrum is reproduced.
Iterations are pursued until all parameter correlations reach their minimum. 
Next, departures from the best-fit model are calculated in one dimension until the corresponding level of confidence is reached for the given number of free parameters and 1$\sigma$ error bars are determined. 
{\sc XTgrid} is fully integrated with the MAIDS cluster and makes use of its multi-processing capabilities.
Another important feature of the iterative steepest-descent fitting method with models calculated on-the-fly is the ability to include the radiative interactions of the components in close binaries \citep{vuckovic16}, which would be more complicated with a classical grid based fitting method. 

The fitting procedure is under constant development. 
Every time a new observational challenge is presented to us, that can be addressed with the theoretical models, the fitting procedure is extended and adjusted to make the most out of the given observation.
There are certain model configurations that have been developed and tested in the past:
\begin{itemize}
\item Multi-wavelength non-LTE spectral fits to hot stars.
\item Binary spectral decomposition with {\sc Tlusty} models + [MILES templates, {\sc Phoenix} or {\sc Atlas} spectra].
\item Binary spectral decomposition with {\sc Tlusty} + {\sc Tlusty} models.
\item Spectral models for hydrogen deficient stars.
\end{itemize}

\subsection{Media contents}
\begin{figure}
\includegraphics[width=\linewidth]{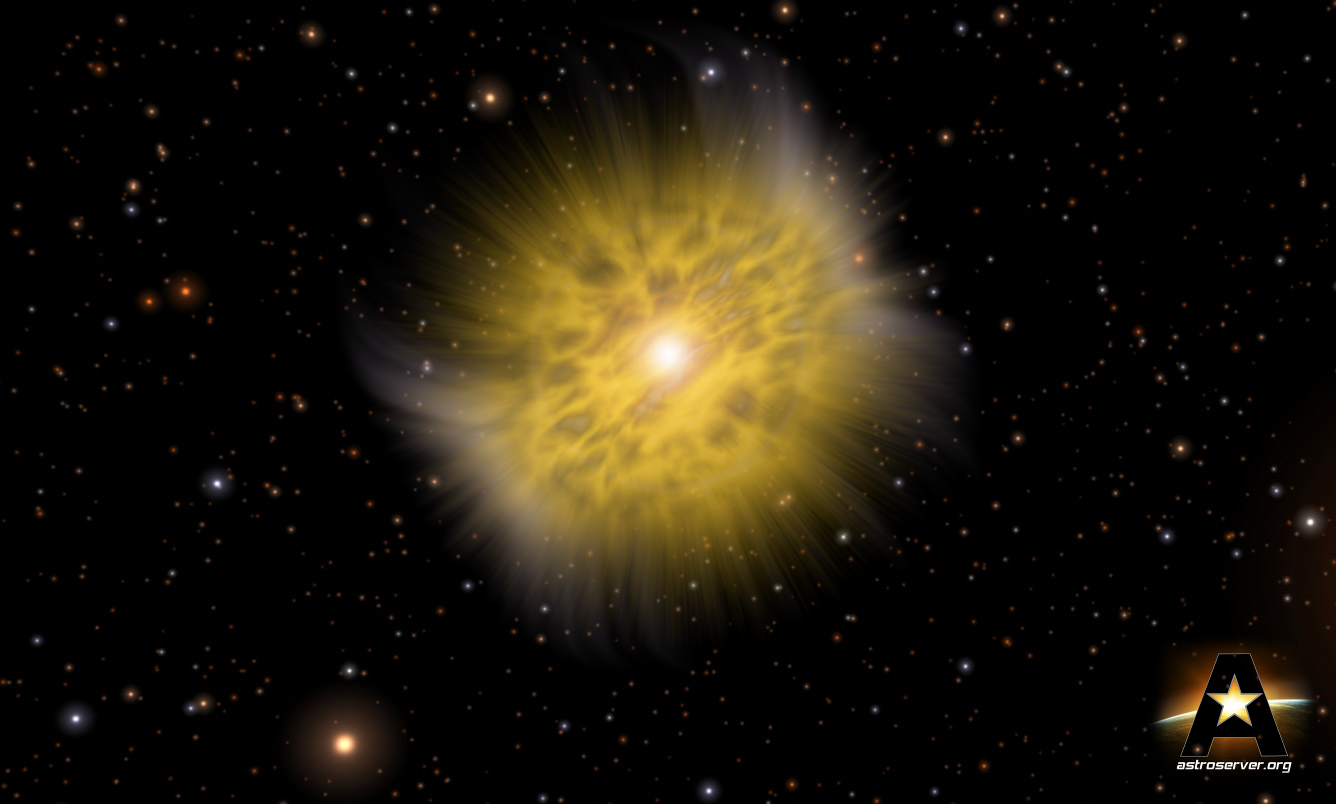}
\caption{Artist's concept of a fictional stellar explosion. (Image credit: Peter Nemeth/Astroserver.org)\label{fig1}}
\end{figure}
Science communication is a cornerstone of any research. 
It is not enough to search for and find the answers to the most exiting natural phenomena, the whole process must be presented in an easily understandable way. 
Peer-reviewed technical research papers provide a high standard and serve as the most reliable sources, but they fall far from this goal. 
As the popularity of online content consumption, the role of social media, and blogs grew, so did the need for well designed and composed artwork. 
Therefore research publications with a significant impact should be accompanied by press releases that provide a scientifically sound summary in lay terms and fine tuned graphics (e.g. Figure \ref{fig1}).
However, the process does not stop here. 
Nowadays whether a major discovery appears in mainstream media or not boils down to a simple question: Is there an animation with it? 
Indeed, clicks and reader attention is driven by catchy titles and, even without knowing, well placed and carefully designed artwork.
Such contents have great importance in grant or observing proposals as well. 
Even though the decision is made based on the technical details of the project, a relatively short animation can communicate the core message of the proposal much faster and help visualizing and understanding the details of the work better.  

The importance of media content and visualizations is evident. 
However, to most researchers this is not the main focus, at least not until the work is done and the paper is accepted. 
By then however, it turns out to be too late. 
Visualization requires a different skill-set than research. 
It is a job by itself. 
Therefore Astroserver teams up with experts in 3D modeling technologies to help speed up such productions. 
No major research results with a public interest should stay under the radar just because it is not supplemented with appropriate media content.

\subsection{Observing}
Visitor-mode observing is a real highlight of astronomy as a profession. 
Traveling to the most exotic and sometimes very distant places on Earth may sound tempting to an adventurous soul. 
However, considering the preparation, scheduling, booking requirements that needs to be done, this idyll quickly disperses. 
In addition, going through instrument manuals and preparing target and backup lists with feasibility constraints makes it nothing more than field work. 
This is how the joyride becomes a mission, after which the successful observer returns home tired with a satisfied happy smile, and the one who had to deal with poor visibility is just tired and happy to be back home. 
Even only a few nights of observing can fill several weeks with tasks related to an upcoming observing run. 
Therefore it is not exaggeration to consider any kind of visitor mode observing to a job by itself. 

This job is perfect for enthusiastic students looking for experience and for young researchers. 
They are flexible, dynamic and independent, able to travel the world and do the great things. 
Remote support usually offers very limited help to observers, however it becomes more important at facilities where there is no local support or operator service.
Astroserver offers night time remote observing support and can take up observing runs.



\section{Astroserver contributions\label{contrib}}
\subsection{Composite spectrum sdB binaries}

Composite spectrum binaries (or multiple systems) are real treasure troves in astronomy. 
One can hit two birds with one stone, as with a single observation one records the spectra of two stars. 
In fact, it is even more than that. 
A follow-up also records the orbit of the binary and reveals the common evolution history and past interactions of the members.
However, this wealth of diagnostics is not readily available, and does not come cheap. 
Hot star spectroscopy is suffering from uncertain atomic data and complicated atmospheric structure with non-LTE effects and, hence, is challenging by itself. 
The situation is not much better on the cool end of the Hertzsprung–Russell diagram. 
The surface gravity diagnostics of cool stars is poor and the occurrence of molecules make the models complicated. 
This is why each field is challenging individually and their joint application for composite spectrum binaries requires more care. 
Even a few percent variation in the flux contribution to the composite spectrum (dilution) or several km\,s$^{-1}$ in the projected rotation velocity is enough to change the derived metallicity and lead to different results. 
The same is true for surface temperature, and these degeneracies together make spectral disentangling a real challenge. 
This is also the reason why the members of composite spectrum binaries cannot, and should not be investigated separately. 
These binaries are complete systems, which should be addressed with a modeling approach that considers the principal components simultaneously.
To find the flux contributions of the members the best spectral range is where the continuum of both members are clearly visible, or where isolated and well calibrated lines of both members are visible. 
The most reassuring thing is, however, when the composite spectrum is reproduced over a large spectral range, including strong and weak lines, multiple ionization stages and several elements of both stellar components. 
This is the direction in which we wish to proceed with the development of {\sc XTgrid}, as demonstrated in Section \ref{sect:UVES} and Figure \ref{fig3}.


\subsubsection{PB 3877 -- a hypervelocity sdB binary}

PB\,3877 has been discovered as a halo sdB star with peculiar kinematics and a high probability that the star is unbound to the Galaxy \citep{tillich11}. 
To investigate the star further new observations have been obtained with the Keck-ESI and VLT-{\sc Xshooter} spectrographs. 
The higher quality Keck-ESI data immediately showed lines of a cool companion superimposed on the sdB spectrum.
The two stars show a very similar radial velocity, therefore a chance alignment can be ruled out and the stars form a binary system.
The infrared excess also confirmed the binarity. 
A spectral decomposition was performed to investigate the binary members, find the distance to the system and refine its kinematics. 
The binary resembles to long period subdwarf binaries that are well known in the disk populations. 
Although such binaries may occur in the halo, they are poorly known due to their large distance. 
A more interesting question is how PB\,3877 obtained its peculiar kinematics. 
While it has a comparable Galactic rest-frame velocity to hypervelocity stars ejected by the super-massive black-hole in the Galactic Center, PB\,3877 never came anywhere near the center. 
Also, any close encounter acceleration mechanism would rip the binary apart. 
Therefore PB\,3877 must be either a bona fide peculiar halo system, or, more likely, it was accreted by the Milky Way from the debris of a disrupted satellite galaxy \citep{nemeth16}.

\begin{figure}
\includegraphics[width=\linewidth]{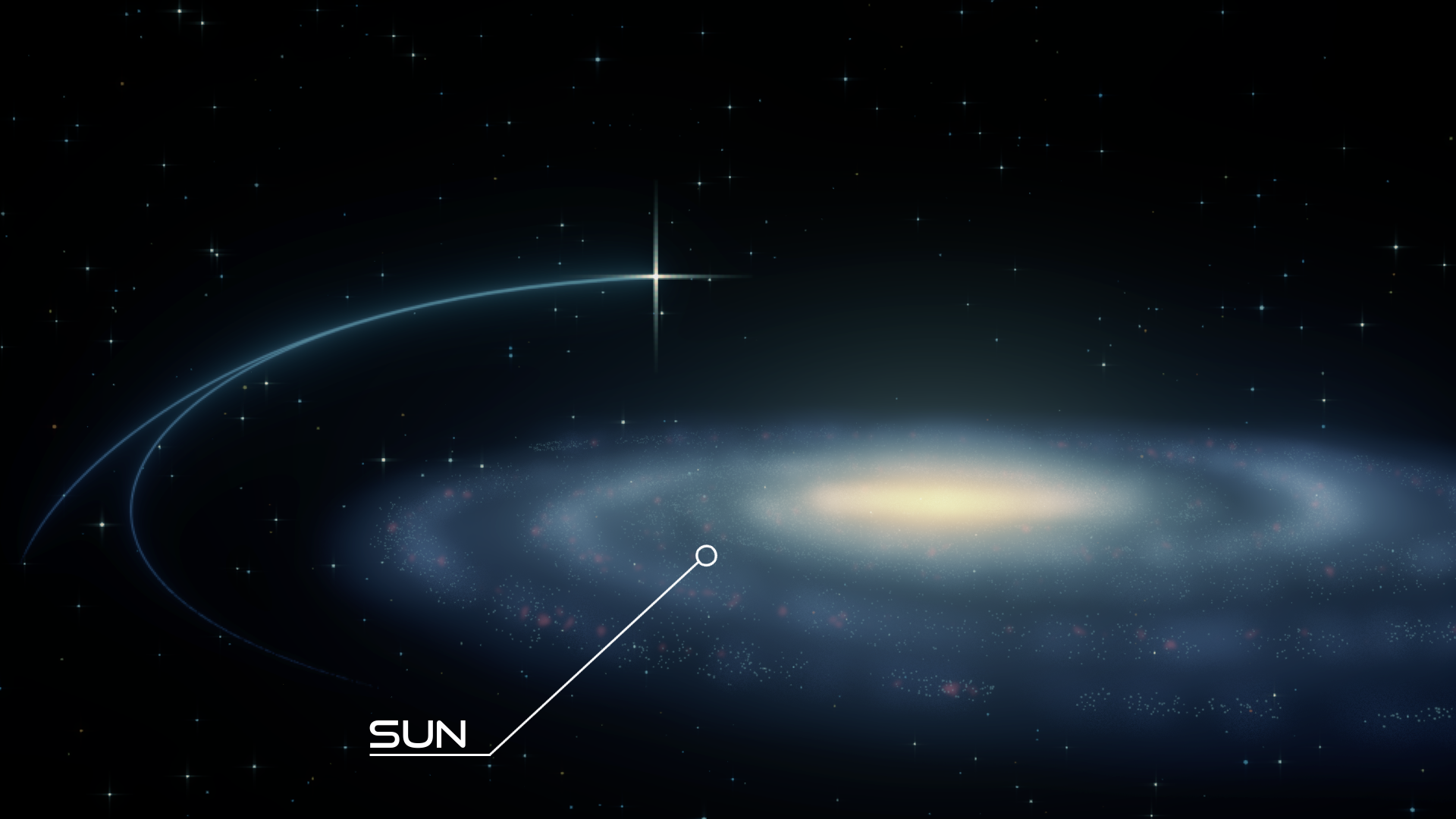}
\caption{PB\,3877 is a hypervelocity wide binary system zooming through the outskirts of the Milky Way galaxy. This artist's concept shows its past trajectory and current location. (Image credit: Thorsten Brand/Astroserver.org) \label{fig2}}
\end{figure}

Astroserver conducted {\sc XTgrid} based model atmosphere analysis of the composite spectrum, made text contributions and took part in the press release artwork production (Figure \ref{fig2}). 

\vspace{5pt}
\noindent More information on PB\,3877 is available at:
\newline\centerline{\url{www.astroserver.org/fxru96}}

\subsubsection{UVES sample\label{sect:UVES}}
\begin{figure*}
\includegraphics[width=\textwidth]{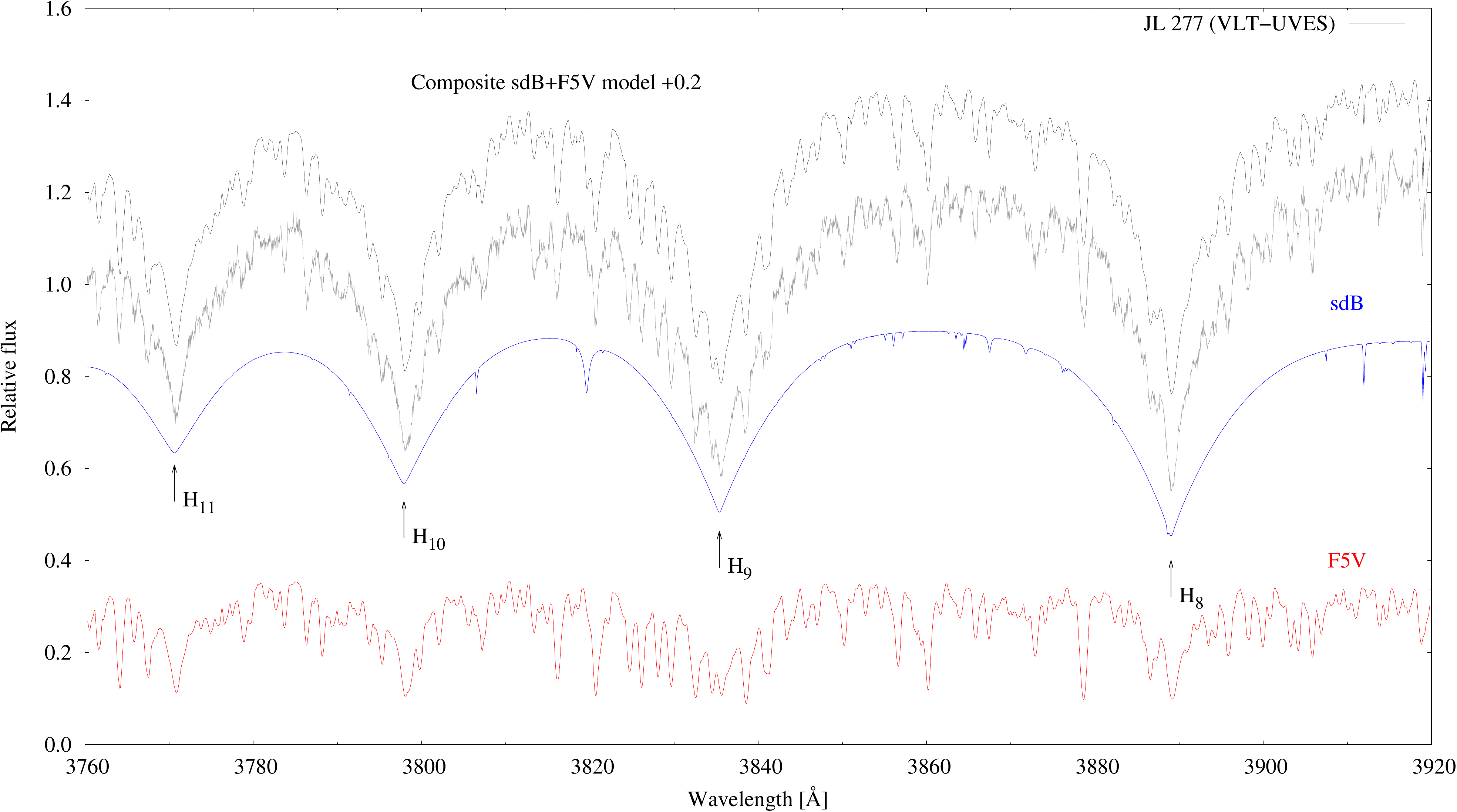}
\vskip 40pt
\includegraphics[width=\textwidth]{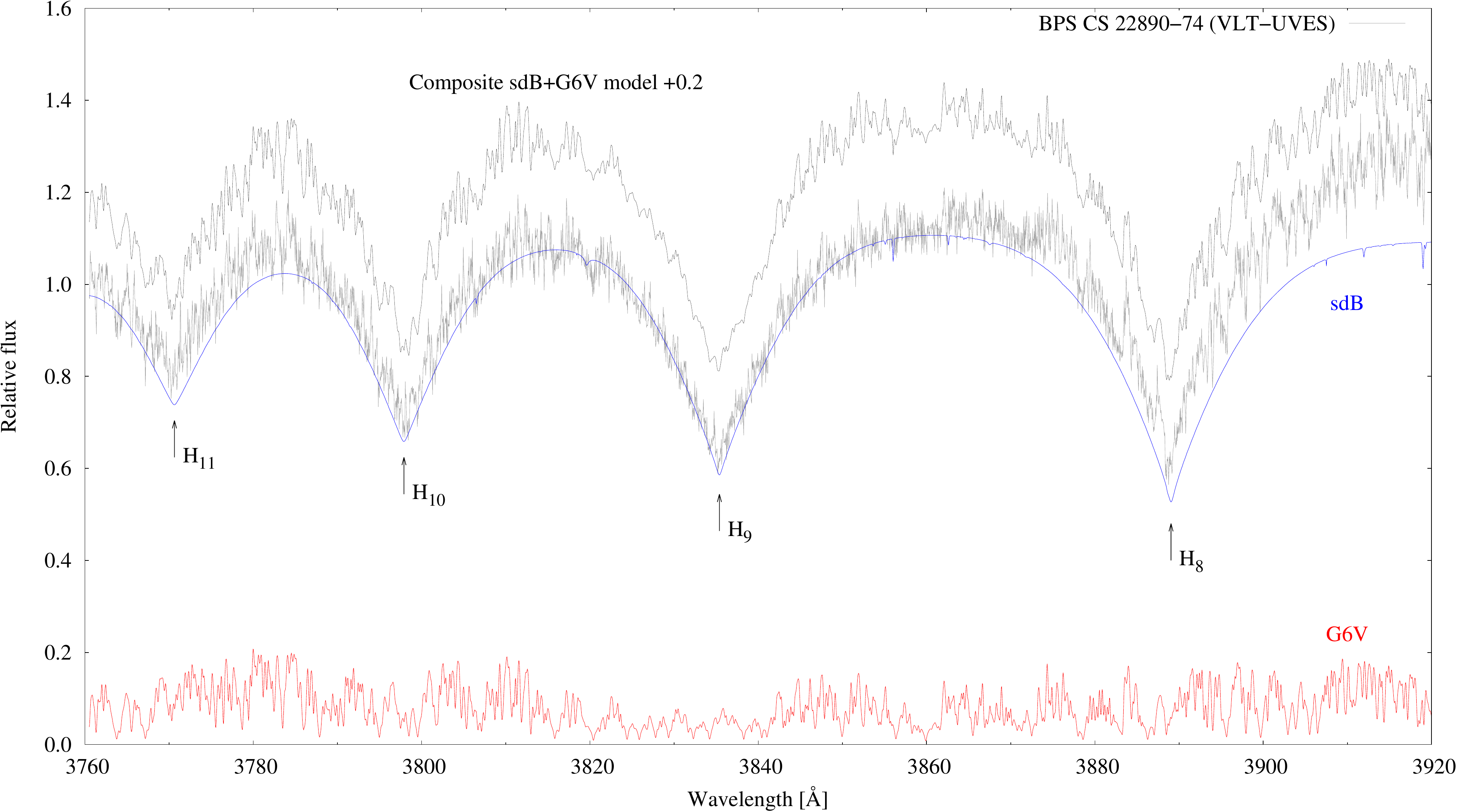}
\caption{Small section from the spectral decomposition of the  sdB+F5V binary JL\,277 (top) and the sdB+G6V binary BPS\,CS\,22890-74 (bottom). The VLT-UVES observation covers the entire optical range from 3760 \AA\ into the near-infrared.\label{fig3}}
\end{figure*}

\cite{vos18} (see also these proceedings) have presented a spectroscopically confirmed subdwarf binary sample.  
The main selection criterion was the brightness of the systems in order to be able to use multiple telescopes in the radial velocity follow-up and obtain a high enough signal-to-noise for abundance analysis.
High resolution optical spectra have been obtained with the VLT-UVES spectrograph on a time-span comparable to the long orbital period (\citealt{chen13} predict $400<P_{\rm orb} <1600$ d) of these binaries. 
The first part of the analysis consists of only nine systems out of the 148 binaries in the full sample. 
However, even this small selection shows a remarkable diversity in the companion types (F, G, K), abundances  
($-1<[{\rm Fe}/{\rm H}]<1$) and projected rotation velocities ($5\ {\rm km\,s}^{-1} <v\sin{i} < 35\ {\rm km\,s}^{-1}$). 
Figure \ref{fig3} shows the spectral decomposition of the sdB+F5V type binary JL\,277 and the sdB+G6V type binary BPS\,CS\,22890-74 in the 3760-3920 \AA\ spectral range. The main properties of these two systems are listed in Table \ref{tab1}.

\begin{table*}[tbp]
\centering
\caption{Atmospheric parameters are shown for the two binary systems in Figure \ref{fig3} from the UVES sample. Subscript 1 refers to the hot subdwarf, for which the models included H, He, C, N, O, Mg, Si and Fe opacities. Subscript 2 refers to the cool companions for which models were extracted from the {\sc Phoenix} spectral library \citep{husser13}. \label{tab1}}
\setstretch{1.3}
\begin{tabular}{l | lcc c | lcrc}
\hline
 $ {\rm System}$            
 & \multicolumn{1}{c}{$T_{\rm eff,1}$}    
 & $\log{g}_1$ 
 & $\log (n{\rm He}/n{\rm H})_1$ 
 & \multicolumn{1}{c|}{${\rm v}\sin{i_1}$} 
 & \multicolumn{1}{c}{$T_{\rm eff,2}$} 
 & $\log{g}_2$ 
 & ${\rm [Fe/H]_2}$          
 & \multicolumn{1}{c}{${\rm v}\sin{i_2}$} \\
 & \multicolumn{1}{c}{${\rm (K)}$}    
 & \multicolumn{1}{c}{${\rm (cm\,s}^{-2}{\rm )}$} 
 & 
 & \multicolumn{1}{c|}{${\rm (km\,s}^{-1}{\rm )}$} 
 & \multicolumn{1}{c}{${\rm (K)}$} 
 & \multicolumn{1}{c}{${\rm (cm\,s}^{-2}{\rm )}$} 
 & 
 & \multicolumn{1}{c}{${\rm (km\,s}^{-1}{\rm )}$} \\
 \cline{2-9}
 
${\rm JL\,277}$           & $25410 ^{+320}_{-940}$ & $5.44^{+0.04}_{-0.23}$     & $-2.48^{+0.10}_{-0.17}$     &  $0$                              & $6270^{+80}_{-130}$ & $4.11^{+0.15}_{-0.15}$     & $-0.5^{+0.1}_{-0.1}$ & $32^{+2.3}_{-1.3}$ \\
${\rm BPS\,CS\,22890-74}$ & $22890^{+1320}_{-320}$ & $5.24^{+0.18}_{-0.07}$     & $-3.03^{+0.60}_{-0.32}$     &  $0$                              & $5650^{+70}_{-70}$  & $4.72^{+0.23}_{-0.26}$     & $\ 0.5^{+0.1}_{-0.1}$  &  $9^{+0.3}_{-0.3}$ \\
\hline
\end{tabular}
\end{table*}

The study concluded that companions of hot subdwarfs accrete only very little mass during the Roche lobe overflow evolution. 

\vspace{5pt}
\noindent More information on composite spectra subdwarf binaries and the fits made to the UVES selection over the entire available spectral range can be found online:
\newline\centerline{\url{www.astroserver.org/kw32yz}}

\subsubsection{FEROS sample}
Another on-going project on composite spectrum binaries by Vennes et al. (see these proceedings) have initiated a small survey of twenty hot subdwarfs using the Fiber-fed Extended Range Optical Spectrograph (FEROS) and the 2.2-m telescope at La Silla. 
The diversity seen in the UVES sample is also present among the FEROS targets, including a long-period (62 d) sdO plus a G III binary (GALEX\,J2038-2657) and several short period binaries with periods ranging from 3.5 hours to 5 days comprised of a hot subdwarf primary with a late-type or a white dwarf companion. 

The orbital period range between 30 and 400 days is particularly interesting from a theoretical aspect of population synthesis. 
These models are the tools to investigate the evolution of binary systems and the formation of hot subdwarfs. 
Common envelope (CE) evolution explains short-period subdwarf binaries with low-mass main sequence or white dwarf companions. 
Plenty of such systems are known with properties matching the predictions nicely. 
On the long-period end, stellar evolution leads to Roche lobe overflow and it explains long-period binaries. 
The long orbital period slows down observing programs and therefore only a handful of systems are solved. 
\citet{vos17} have investigated such binaries and found a positive correlation between orbital period and eccentricity, which makes long-period binaries important in fine-tuning population synthesis models. 
A very compelling region is in between the two period regimes, which is almost empty. 
Population synthesis predicts sdB+A systems that are not observed among the nearly 200 solved subdwarf binaries. 
This is why systems in the gap, such as GALEX\,J2038-2657, are particularly valuable for learning about binary evolution. 
Despite the facts that the sdO star in GALEX\,J2038-2657 is likely more massive than 0.5 $M_\odot$, it avoided the extreme horizontal branch and now evolves toward the white dwarf cooling sequence. 
The system is in between two CE phases and the companion is in a transient evolutionary phase, such as the Hertzsprung-gap or the early giant-branch. 
With a hot subdwarf lifetime of $\sim$100 Myrs and a subgiant companion lifetime of a few Myrs such binaries are quite rare. 
\\
\citet{moni15} found an sdB star (\#5865) with a main sequence K type companion on a 1.61 d orbit in the globular cluster NGC\,6752. 
This object represents a new class of sdB binaries not previously observed in the Galactic field. 
The system is analogous to the short-period sdB+FGK binaries predicted by population synthesis. 
Although the lifetime of main sequence K stars is much longer than the sdB lifetime, the evolutionary stage of \#5865 is similar to that of GALEX\,J2038-2657. 
However, the very short orbital period of \#5865 is substantially different, and it may be related to the crowded and dynamic environment in the globular cluster as well as to the lower mass of the companion. 
Future work is needed to investigate these interesting binaries. 

\vspace{5pt}
\noindent More information on composite spectra subdwarf binaries and the FEROS sample is available online:
\newline\centerline{\url{www.astroserver.org/vpd0aj}}


\subsection{LP 40-365 -- the first SN Iax remnant candidate}

\cite{vennes17} have observed LP\,40-365 with the 4-m Mayall telescope and RC spectrograph as a bright backup target as they were unable to conduct their main science program due to poor weather conditions. 
The star was selected kinematically from the Luyten-Palomar survey as it is relatively bright and shows a large proper motion. 
The first spectra revealed a cool object dominated by singly ionized metals, with excess surface gravity and missing hydrogen lines.
The heliocentric radial velocity was found to be close to 500 km\,s$^{-1}$, which is only one component of the total velocity, but already close to the Galactic escape-velocity. 
The surface gravity suggested a low-mass compact object, but the temperature corresponded to an old white dwarf. 
However, the most intriguing property of LP\,40-365 was its metal-rich surface composition. 
These observations made LP\,40-365 a top-priority target for follow-up. 
The subsequent observations with WHT-ISIS, MDM-Modspec and Gemini-GRACES revealed that LP\,40-365 is most likely a hypervelocity low-mass compact remnant of a failed supernova Iax explosion. 

\begin{figure}
\includegraphics[width=\linewidth]{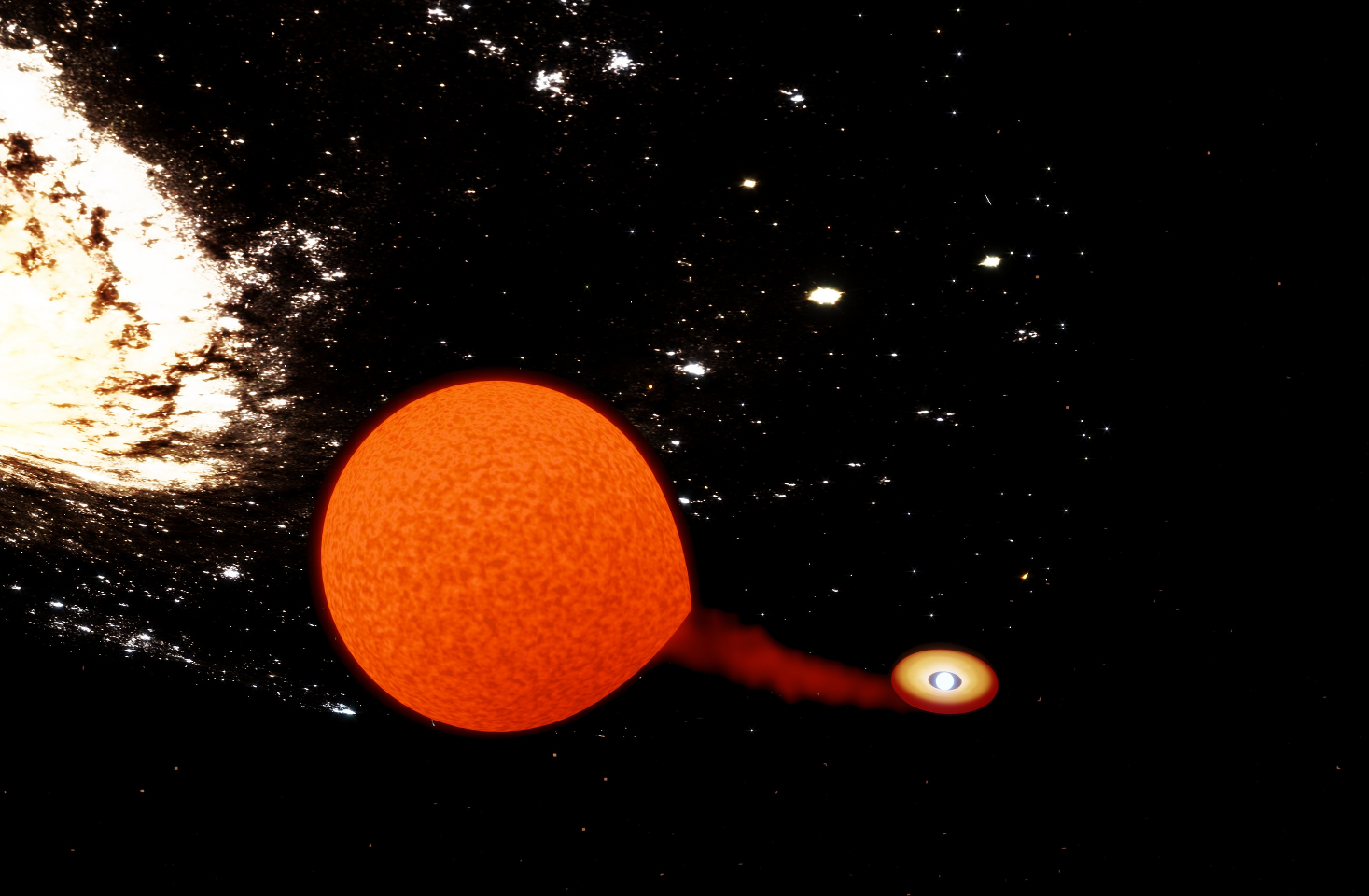}
\caption{Artist's impression of the progenitor binary system of LP\,40-365 far out in the Galactic halo. (Image credit: Sardi Pax/Astroserver.org) \label{fig4}}
\end{figure}

Astroserver took part in the observing proposal preparation, conducted follow-up observations, {\sc XTgrid} model atmosphere analysis and coordinated part of the press release image (Figure \ref{fig4}) and animation production.

\vspace{5pt}
\noindent More information on LP\,40-365 is available online:
\newline\centerline{ \url{www.astroserver.org/nwkzka}}

\section{Summary\label{summary}}

We have made an attempt to introduce the current socio-cultural drifts in research astronomy.
These shifts show that professional research can be subdivided to several tasks and each of these tasks may represent a full job.
Independent researchers (freelancers) may contribute efficiently to specific tasks in large projects.
We have also introduced Astroserver, a new initiative for freelancers.
It is a non-profit organization which provides computing and observing services as well as a framework for collaborative research. 
Interested professionals are encouraged to consider these services in their research projects or educational program.
Any feedback is appreciated on expertise, either offered or searched for in this framework, including experience with observing techniques, instruments, data reduction, analysis and processing, specialised code and algorithms.  
\\
One cornerstone of the currently offered services by Astroserver is model atmosphere work for hot evolved stars and binaries. 
Beyond our on-going work on atomic diffusion in the atmosphere of the blue horizontal branch star Feige\,86 (these proceedings), we presented further examples from recent projects with Astroserver contributions. 
These projects focused on small surveys of composite spectrum subdwarf binaries observed with UVES and FEROS, and found an unexpected diversity among the observed systems. 
Another set of projects dealt with objects showing a peculiar composition or kinematics. 
PB\,3877 turned out to be a halo sdB+K wide binary and LP\,40-365 is a low-mass white dwarf, and very likely a bound remnant of a failed supernova Iax. 
On top of these peculiarities, both objects were found to be hypervelocity systems. 
\\
Astroserver is accepting projects.
However, without external funding the given running costs of MAIDS and the FTE requirements do not allow these services to be offered for free. 
On the other hand, to promote future collaborations and help the corresponding funding, grant and observing proposals, we provide pilot-studies and feasibility tests for free, if Astroserver is an active participant in the project.
\\
We encourage interested researchers to browse the Astroserver web site and provide feedback:
\newline\centerline{\url{www.astroserver.org/survey}}

\end{document}